\documentclass[11pt]{article} 
\usepackage{fullpage}        
\usepackage{setspace}        
\usepackage{float}           
\usepackage{placeins}         

\usepackage{kotex}            
\usepackage[utf8]{inputenc}  

\usepackage{amsmath}          
\usepackage{amsfonts}         
\usepackage{amssymb}          
\usepackage{amsthm}           
\usepackage{mathtools}        

\usepackage{booktabs}         
\usepackage{caption}
\usepackage{subcaption}
\usepackage[round]{natbib}    

\usepackage{algorithm}        
\usepackage{algpseudocode}    

\usepackage[dvipsnames]{xcolor}
\usepackage[colorlinks=true,  
            linkcolor=blue,
            urlcolor=blue,
            citecolor=blue]{hyperref} 

\newtheorem{remark}{Remark}

\title{On parameter estimation for the truncated skew-normal distribution}
\date{}

\author{
Kwangok Seo\thanks{Department of Statistics, Inha University, Incheon, Korea.}
\and
Seul Lee\thanks{Department of Statistics, Seoul National University, Seoul, Korea.}
\and
Johan Lim\footnotemark[2]
\thanks{Corresponding author. Email: johanlim@snu.ac.kr}
}

\begin{document}

\maketitle 

\begin{abstract}
Parameter estimation for the truncated skew-normal distribution is challenging, as truncation introduces additional nonlinearity into the likelihood function and often leads to numerical instability in existing estimation procedures. In this paper, we propose a grid-based estimation method, referred to as GRID-MOM, for parameter estimation in the truncated skew-normal distribution. The proposed approach fixes the shape parameter on a pre-specified grid and, for each grid point, estimates the location and scale parameters using the method of moments. The optimal value of the shape parameter is then selected via likelihood-based comparison, yielding the final parameter estimates. By decoupling the estimation of the shape parameter from that of the location and scale parameters, the proposed method reduces the complexity of the optimization problem and improves numerical stability. We evaluate the finite-sample performance of the proposed estimator through an extensive numerical study, comparing it with existing methods under a variety of scenarios. The results demonstrate that the proposed method provides stable and accurate estimation, particularly for the shape parameter, suggesting that the proposed method offers a practical alternative for inference in truncated skew-normal models. We further demonstrate the practical applicability of the proposed method using phosphoproteomics data and hospital admission data.

\end{abstract}

\section{Introduction}
The skew-normal distribution provides a flexible extension of the normal distribution by accommodating asymmetry through an additional shape parameter, while retaining much of the analytical tractability of the Gaussian model \citep{azzalini2013skew}. Since its introduction, the skew-normal family has been widely used in statistical modeling across a broad range of applications.

In many practical settings, observations are subject to truncation. Such truncation naturally arises in contexts including reliability analysis with detection limits, biomedical and environmental studies involving physical or regulatory bounds, and socio-economic data subject to reporting thresholds. Data generated in these settings also frequently exhibit substantial asymmetry, a feature that cannot be adequately captured by symmetric truncated models. In such cases, the truncated skew-normal distribution provides a natural and flexible modeling framework by jointly accommodating truncation and skewness within a single parametric family \citep{kim2004family}.

Several approaches have been proposed for parameter estimation in the truncated skew-normal distribution. Maximum likelihood estimation (MLE) is a natural choice, as it directly exploits the likelihood structure of the truncated model. Alternatively, moment-based methods estimate the parameters by matching empirical (weighted) moments to their theoretical counterparts. For instance, the method of moments (MOM; \citealp{martinez2008note, flecher2010truncated}) relies on the first three moments, whereas the method of weighted moments (MWM; \citealp{flecher2010truncated}) extends MOM by incorporating suitably constructed weighted moments to enhance estimation stability.

Despite the availability of these methods, achieving stable and accurate estimation remains challenging, particularly in the presence of pronounced skewness and substantial truncation. These observations motivate the development of alternative estimation procedures.

In this paper, we propose a new estimation methodology for the truncated skew-normal distribution. The proposed approach is designed to improve numerical stability while remaining computationally efficient and straightforward to implement. Its performance is assessed through an extensive numerical study. The results indicate that the proposed method offers favorable finite-sample performance and enhanced numerical stability across a range of practically relevant truncation scenarios.

The remainder of the paper is organized as follows. Section~\ref{sec:preliminaries} reviews the truncated skew-normal distribution and summarizes existing estimation methods. Section~\ref{sec:method} introduces the proposed estimation procedure and discusses its implementation details. Section~\ref{sec:simulation} presents the numerical study and comparative results. Section~\ref{sec:realdata} illustrates the practical applicability of the proposed method through analyses of phosphoproteomics data and hospital admission data.  Section~\ref{sec:discussion} concludes with a summary.

\section{Preliminaries}\label{sec:preliminaries}
We begin by reviewing the skew-normal and truncated skew-normal distributions.

\paragraph{Skew-normal distribution}
Let $X$ be a univariate skew-normal random variable with location parameter $\xi \in \mathbb{R}$, scale parameter $\omega > 0$, and shape parameter $\alpha \in \mathbb{R}$. Following the formulation of \citet{azzalini1985class}, the density function of $X$ is given by
\begin{equation}\label{eq:sn_pdf}
    f_{\mathrm{SN}}(x \mid \xi, \omega, \alpha)
    = \frac{2}{\omega}
    \phi\!\left( \frac{x - \xi}{\omega} \right)
    \Phi\!\left( \alpha \frac{x - \xi}{\omega} \right),
\end{equation}
where $\phi(\cdot)$ and $\Phi(\cdot)$ denote the standard normal density and distribution functions, respectively. The skew-normal distribution provides a extension of the normal model by allowing for skewness through the shape parameter $\alpha$. When $\alpha = 0$, the skew-normal distribution reduces to the normal distribution with mean $\xi$ and variance $\omega^2$. 

\paragraph{Truncated skew-normal distribution}
Suppose that the random variable $ X $ follows a skew-normal distribution as in \eqref{eq:sn_pdf}, but is observed only within the truncation interval $ [L, U] $, where $ -\infty \le L < U \le \infty $. In this case, the conditional distribution of $ X $ given $ X \in [L, U] $ is the truncated skew-normal distribution, whose density is
\begin{equation}\label{eq:tsn_pdf}
    f_{\mathrm{TSN}}(x \mid \xi, \omega, \alpha)
    =
    \frac{
    f_{\mathrm{SN}}(x \mid \xi, \omega, \alpha)
    }{
    F_{\mathrm{SN}}(U \mid \xi, \omega, \alpha)
    -
    F_{\mathrm{SN}}(L \mid \xi, \omega, \alpha)
    }
    \mathbb{I}(L \le x \le U),
\end{equation}
where $ F_{\mathrm{SN}}(\cdot \mid \xi, \omega, \alpha) $ denotes the distribution function of the skew-normal distribution. Throughout, we treat the truncation interval $ [L, U] $ as known and focus on estimating the parameters $ \xi $, $ \omega $, and $ \alpha $ from the truncated observations.

\bigskip
We next review representative parameter estimation methods for the truncated skew-normal distribution, including maximum likelihood estimation (MLE), the method of moments (MOM), and the method of weighted moments (MWM). Let $ X_1, \ldots, X_n $ be independent and identically distributed observations from the truncated skew-normal distribution defined in \eqref{eq:tsn_pdf}, and let $ x_1, \ldots, x_n $ denote their realizations.

\paragraph{Maximum likelihood estimation}
The log-likelihood function for the truncated skew-normal distribution can be written as
\begin{equation}\label{eq:tsn_loglik}
    \ell_{\mathrm{TSN}}(\xi, \omega, \alpha)
    =
    \sum_{i=1}^n
    \log f_{\mathrm{SN}}(x_i \mid \xi, \omega, \alpha)
    -
    n \log
    \left\{
    F_{\mathrm{SN}}(U \mid \xi, \omega, \alpha)
    -
    F_{\mathrm{SN}}(L \mid \xi, \omega, \alpha)
    \right\}.
\end{equation}
The MLE is obtained by maximizing the log-likelihood function \eqref{eq:tsn_loglik} with respect to $(\xi, \omega, \alpha)$. In contrast to the untruncated case, the likelihood function for the truncated model depends on all parameters through the normalization constant, making the optimization problem substantially more complex. Since the log-likelihood in \eqref{eq:tsn_loglik} does not admit a closed-form solution, the maximum likelihood estimator must be obtained via numerical optimization.

\paragraph{Method of moments} The MOM estimates the model parameters by equating empirical moments to their theoretical counterparts under the truncated skew-normal distribution. Specifically, MOM estimators are obtained by solving a system of equations of the form
\begin{equation*}
    \mathbb{E}_{\mathrm{TSN}}(X^k \mid L \leq X \leq U) = \frac{1}{n} \sum_{i=1}^n x_i^k,
    \quad k = 1, 2, 3,    
\end{equation*}
where $\mathbb{E}_{\mathrm{TSN}}(\cdot \mid L \leq X \leq U)$ denotes the expectation with respect to the truncated skew-normal distribution, conditional on the truncation interval $[L, U]$. Proposition~2 of \cite{flecher2010truncated} provides closed-form expressions for the first three theoretical moments, which facilitate the implementation of the MOM.

\paragraph{Method of weighted moments}
The MWM, originally proposed for the skew-normal distribution by \cite{flecher2009estimating}, was extended to the truncated skew-normal distribution by \cite{flecher2010truncated} to improve the performance of the method of moments. This method is based on weighted moments of the form
\begin{equation}\label{eq:wm}
    \mathbb{E}\left[X^k \Phi^r(X) \mid L \leq X \leq U\right].
\end{equation}
When $ r = 0 $, the weighted moments in \eqref{eq:wm} reduce to the ordinary moments. Specifically, the MWM estimator proposed by \cite{flecher2010truncated} is obtained by solving the following system of equations:
\begin{align*}
    \mathbb{E}_{\mathrm{TSN}}\left[X \mid L \leq X \leq U\right] 
    &= \frac{1}{n}\sum_{i = 1}^n x_i \eqqcolon \bar{x},\\
    \mathrm{Var}_{\mathrm{TSN}}(X \mid L \leq X \leq U) 
    &= \frac{1}{n} \sum_{i = 1}^n (x_i - \bar{x})^2 \eqqcolon s^2,\\
    \mathbb{E}_{\mathrm{TSN}}\left[\Phi(X) \mid L \leq X \leq U \right] 
    &= \frac{1}{n} \sum_{i = 1}^n \Phi(x_i).
\end{align*}
Closed-form expressions for the theoretical quantities appearing on the left-hand sides of the above equations can be derived from Propositions~2 and~3 of \cite{flecher2010truncated}, which facilitate the implementation of the MWM.

\bigskip
We conclude this section by summarizing the limitations of the existing methods.

\paragraph{Limitations of existing methods.}
MLE estimates the parameters by maximizing the log-likelihood in \eqref{eq:tsn_loglik}. Because the log-likelihood is generally nonconcave in $(\xi,\omega,\alpha)$, the optimization problem may admit multiple local maximizers. Consequently, standard numerical algorithms can converge to suboptimal solutions, rendering the estimates sensitive to initialization.
MOM estimates the parameters by matching the first three moments of the truncated skew-normal distribution. The third moment has a more complex form and typically exhibits greater variability than the first two, which may induce numerical instability in finite samples.
MWM replaces the third moment with a suitably constructed weighted moment to improve stability. Although this modification often enhances numerical performance, a fundamental limitation remains. When the shape parameter is large (e.g., $\alpha \ge 4$), the weighted moment varies only marginally with further increases in $\alpha$, providing limited information for distinguishing among large values. As a result, accurate estimation under pronounced skewness remains difficult.
\section{Method}\label{sec:method}
Up to this point, we have introduced the truncated skew-normal distribution and reviewed three existing estimation methods. We now describe the proposed method. The key idea of the proposed method is to decouple the estimation of the shape parameter from that of the location and scale parameters, thereby reducing the complexity of the optimization problem and improving numerical stability. 

Let $\mathcal{G} = \{\alpha_1, \ldots, \alpha_G\}$ denote a prespecified grid of candidate values for the shape parameter. For each fixed $\alpha_g \in \mathcal{G}$, the location and scale parameters $(\xi, \omega)$ are estimated conditionally on $\alpha = \alpha_g$ via the method of moments under the truncated skew-normal distribution. Specifically, given $\alpha = \alpha_g$, we solve
\begin{align*}
    \mathbb{E}_{\mathrm{TSN}}\!\left[X \mid \alpha = \alpha_g, L \le X \le U \right] &= \bar{x}, \\
    \mathrm{Var}_{\mathrm{TSN}}\!\left(X \mid \alpha = \alpha_g, L \le X \le U \right) &= s^2. 
\end{align*}
These equations involve only $(\xi,\omega)$ and can be solved numerically using standard root-finding algorithms. Let $(\hat{\xi}(\alpha_g), \hat{\omega}(\alpha_g))$ denote the resulting solution.
For each $\alpha_g \in \mathcal{G}$, we then evaluate the truncated skew-normal log-likelihood
\begin{equation*}
    \ell_{\mathrm{TSN}}\bigl(\hat{\xi}(\alpha_g), \hat{\omega}(\alpha_g), \alpha_g\bigr).
\end{equation*}
The final estimator is the triplet $(\hat{\xi}, \hat{\omega}, \hat{\alpha})$, where
\begin{equation*}
    \hat{\alpha}
    =
    \arg\max_{\alpha_g \in \mathcal{G}}
    \ \ell_{\mathrm{TSN}}\bigl(\hat{\xi}(\alpha_g), \hat{\omega}(\alpha_g), \alpha_g\bigr),    
\end{equation*}
and $\hat{\xi} = \hat{\xi}(\hat{\alpha})$, $\hat{\omega} = \hat{\omega}(\hat{\alpha})$.

\begin{remark}
    The numerical stability of the proposed method arises from three structural features. 
    First, by conditioning on the shape parameter, the original three-dimensional optimization problem is decomposed into a sequence of two-dimensional subproblems. 
    Second, this formulation avoids reliance on third-order or weighted moments, which are known to be complex and highly variable, particularly in finite samples. 
    Finally, the grid search over the shape parameter, akin to multiple initializations in nonconvex optimization, reduces the risk of convergence to local maximizers.
\end{remark}

We conclude this section with several implementation details of the proposed method. A key component of the procedure is the specification of the grid $\mathcal{G}$ for the shape parameter $\alpha$. In principle, the appropriate range of $\alpha$ depends on the distributional characteristics of the observed data, particularly the degree of skewness. However, in the presence of truncation, the information contained in the tails of the distribution---where the shape parameter $\alpha$ is primarily identified---is partially lost. As a result, reliably determining a reasonable range for $\alpha$ from the data is challenging, and we therefore adopt a heuristic but practical approach. Specifically, we consider a symmetric interval $[-a, a]$ for $\alpha$, with $a > 0$, and construct an equally spaced grid over this interval. That is,
\begin{equation*}
\mathcal{G} = \left\{ -a + (g-1)\Delta : g = 1, \ldots, G \right\},
\qquad
\Delta = \frac{2a}{G-1}.
\end{equation*}
We recommend setting $a = 5$ with $G > 100$, as the range $|\alpha| \leq 5$ is sufficient to cover virtually all levels of skewness encountered in practical applications. If stronger skewness is anticipated, practitioners may expand the range accordingly.

\section{Numerical Study}\label{sec:simulation}
In this section, we conduct simulation studies to evaluate the finite-sample performance of the proposed method, referred to as GRID-MOM. 
Section~\ref{sim_1} compares GRID-MOM with three existing methods---MLE, MOM, and MWM, while Section~\ref{sim_2} compares GRID-MOM with GRID-MLE, a profile-likelihood-based variant (described below).
Throughout the simulation studies, the estimates for all methods are obtained by solving the corresponding optimization problems using a single initial value; multiple initializations are not considered.

\subsection{Comparison with MLE, MOM, and MWM}\label{sim_1}
We adopt the simulation setting of \citet{flecher2010truncated}. Data are generated from a truncated skew-normal distribution with $\xi_0 = 0$, $\omega_0 = 1$, and $\alpha_0 \in \{1, 2, 4\}$ to assess the effect of skewness. 
Three truncation directions are investigated: lower truncation $[L, \infty)$, upper truncation $(-\infty, U]$, and double truncation $[L, U]$. For each truncation direction, we control the truncation rate $\tau \in \{0.1, 0.2\}$. To be specific, for lower truncation, the boundary is set to $L = F^{-1}_{\mathrm{SN}}(\tau \mid \xi_0, \omega_0, \alpha_0)$, while for upper truncation we set $U = F^{-1}_{\mathrm{SN}}(1 - \tau \mid \xi_0, \omega_0, \alpha_0)$. For double truncation, the boundaries are chosen as $L = F^{-1}_{\mathrm{SN}}(\tau/2 \mid \xi_0, \omega_0, \alpha_0)$ and $U = F^{-1}_{\mathrm{SN}}(1 - \tau/2 \mid \xi_0, \omega_0, \alpha_0)$. 
For each configuration, we generate $n = 500$ independent observations from the corresponding truncated skew-normal distribution, and each experiment is replicated $1{,}000$ times. The performance of each estimator is evaluated using multiple summary measures, including empirical bias, root mean squared error (RMSE), the median of the parameter estimates, and the interquartile range (IQR).
The MLE is implemented using the \texttt{R} package \texttt{TSMNSN}. 
The MOM estimator is obtained by solving the corresponding system of equations via the \texttt{R} function \texttt{nleqslv}, while the MWM estimator is computed by minimizing the associated system of equations using the \texttt{R} function \texttt{nlminb}. Tables~\ref{tab:table1}–\ref{tab:table4} summarize the simulation results. 

\paragraph{Results for truncation rate $\tau = 0.1$}
Table~\ref{tab:table1} reports the empirical bias and RMSE of the competing estimators, from which several observations emerge. 
First, consistent with \citet{flecher2010truncated}, MWM generally outperforms MOM across most simulation settings in the sense that it exhibits comparable or smaller bias and RMSE. Accordingly, subsequent comparisons focus on MLE, MWM, and GRID-MOM.
Second, estimation accuracy depends strongly on the truncation direction. All three methods (MLE, MWM and GRID-MOM) perform well under right truncation, whereas left and double truncation result in substantially larger bias and RMSE. 
Third, when $\alpha_0 = 4$ under left truncation, the bias and RMSE of the MLE for the shape parameter exceed 100. This inflation is driven by a subset of replications in which the MLE produces unusually large estimates; among 1{,}000 replications, 150 estimates exceed 10 and 6 exceed 100. 
Finally, when $\alpha_0 = 1$, MLE and MWM exhibit satisfactory performance and outperform GRID-MOM in terms of bias and RMSE. In contrast, for larger shape parameters ($\alpha_0 \in \{2,4\}$), GRID-MOM provides superior performance, particularly in estimating $\alpha$. 

To reduce the influence of extreme values, Table~\ref{tab:table3} reports the median and IQR of the parameter estimates. We observe the following patterns. 
First, consistent with Table~\ref{tab:table1}, MWM generally dominates MOM; across most settings, the median of MWM is closer to the true parameter values, and when the two methods yield similar medians, MWM typically exhibits a comparable or smaller IQR. Accordingly, subsequent comparison focus on MLE, MWM, and GRID-MOM.
Second, under right truncation with $\alpha_0 = 1$, MLE and MWM perform similarly and outperform GRID-MOM; although the median of GRID-MOM remains close to the true parameter values, its IQR for $\hat{\xi}$ and $\hat{\alpha}$ is slightly larger. For larger shape parameters ($\alpha_0 \in \{2,4\}$) under right truncation, all three methods (MLE, MWM and GRID-MOM) yield medians close to the true values with comparable IQRs.
Finally, when the truncation direction is left or double, performance varies with the level of skewness. When $\alpha_0 = 1$, MLE exhibits the best performance. For $\alpha_0 = 2$, MLE and GRID-MOM perform comparably and both outperform MWM; in particular, the median of the MWM estimator for the shape parameter deviates from the true value of 2. When $\alpha_0 = 4$, GRID-MOM outperforms both MLE and MWM.

\begin{table}[!htb]
\centering
\caption{Bias and RMSE of parameter estimates for MLE, MOM, MWM, and GRID-MOM based on $1{,}000$ replications with sample size $n = 500$ under truncation rate $\tau = 0.1$. The true parameters are $\xi_0 = 0$, $\omega_0 = 1$, and $\alpha_0 \in \{1, 2, 4\}$.}
\label{tab:table1}
\begin{tabular}{llcccccccc}
\toprule
Truncation &  & \multicolumn{2}{c}{MLE} & \multicolumn{2}{c}{MOM} & \multicolumn{2}{c}{MWM} & \multicolumn{2}{c}{GRID-MOM} \\
\cmidrule(lr){3-4} \cmidrule(lr){5-6} \cmidrule(lr){7-8} \cmidrule(lr){9-10}
 &  & Bias & RMSE & Bias & RMSE & Bias & RMSE & Bias & RMSE \\
\midrule
\multicolumn{10}{l}{$\alpha_0 = 1$} \\
Left & $\hat{\xi}$ & 0.146 & 0.336 & 0.445 & 0.668 & 0.073 & 0.133 & 0.314 & 0.629 \\
 & $\hat{\omega}$ & -0.018 & 0.144 & 0.081 & 1.552 & -0.033 & 0.070 & 0.075 & 0.222 \\
 & $\hat{\alpha}$ & -0.121 & 0.860 & -0.869 & 2.840 & -0.185 & 0.358 & -0.465 & 1.391 \\
\addlinespace
Right & $\hat{\xi}$ & 0.102 & 0.371 & 0.230 & 0.576 & 0.087 & 0.339 & 0.228 & 0.576 \\
 & $\hat{\omega}$ & 0.053 & 0.285 & 0.085 & 0.262 & 0.049 & 0.281 & 0.088 & 0.271 \\
 & $\hat{\alpha}$ & -0.011 & 0.965 & -0.274 & 1.292 & 0.014 & 0.885 & -0.268 & 1.311 \\
\addlinespace
Double & $\hat{\xi}$ & 0.140 & 0.411 & 0.381 & 0.737 & 0.088 & 0.302 & 0.358 & 0.739 \\
 & $\hat{\omega}$ & 0.029 & 0.214 & 0.111 & 0.218 & 0.013 & 0.197 & 0.131 & 0.226 \\
 & $\hat{\alpha}$ & 0.011 & 1.144 & -0.543 & 1.751 & 0.053 & 0.908 & -0.484 & 1.768 \\
\addlinespace
\midrule
\multicolumn{10}{l}{$\alpha_0 = 2$} \\
Left & $\hat{\xi}$ & 0.058 & 0.172 & 0.392 & 0.503 & 0.079 & 0.164 & 0.090 & 0.301 \\
 & $\hat{\omega}$ & -0.019 & 0.082 & -0.078 & 0.157 & -0.041 & 0.094 & 0.005 & 0.109 \\
 & $\hat{\alpha}$ & 0.209 & 1.563 & $>$100 & $>$100 & -0.490 & 0.961 & 0.041 & 1.342 \\
\addlinespace
Right & $\hat{\xi}$ & 0.028 & 0.143 & 0.036 & 0.171 & 0.039 & 0.147 & 0.032 & 0.167 \\
 & $\hat{\omega}$ & -0.007 & 0.176 & -0.010 & 0.172 & -0.020 & 0.176 & -0.004 & 0.174 \\
 & $\hat{\alpha}$ & 0.013 & 0.760 & -0.016 & 0.793 & -0.058 & 0.746 & 0.005 & 0.792 \\
\addlinespace
Double & $\hat{\xi}$ & 0.071 & 0.228 & 0.291 & 0.467 & 0.099 & 0.180 & 0.114 & 0.363 \\
 & $\hat{\omega}$ & -0.016 & 0.149 & -0.088 & 0.185 & -0.068 & 0.143 & 0.008 & 0.134 \\
 & $\hat{\alpha}$ & 0.082 & 1.296 & $>$100 & $>$100 & -0.440 & 0.958 & -0.053 & 1.420 \\
\addlinespace
\midrule
\multicolumn{10}{l}{$\alpha_0 = 4$} \\
Left & $\hat{\xi}$ & 0.059 & 0.166 & 0.255 & 0.438 & -0.042 & 0.164 & 0.082 & 0.356 \\
 & $\hat{\omega}$ & -0.025 & 0.193 & 0.130 & 1.820 & -0.018 & 0.080 & 0.107 & 0.488 \\
 & $\hat{\alpha}$ & $>$100 & $>$100 & $>$100 & $>$100 & -2.517 & 2.679 & -0.986 & 2.515 \\
\addlinespace
Right & $\hat{\xi}$ & 0.002 & 0.045 & 0.001 & 0.054 & 0.011 & 0.064 & 0.003 & 0.043 \\
 & $\hat{\omega}$ & 0.006 & 0.103 & 0.009 & 0.119 & -0.005 & 0.118 & 0.004 & 0.096 \\
 & $\hat{\alpha}$ & 0.165 & 0.967 & $>$100 & $>$100 & 0.034 & 1.180 & 0.053 & 0.737 \\
\addlinespace
Double & $\hat{\xi}$ & 0.036 & 0.132 & 0.320 & 0.439 & 0.122 & 0.157 & 0.043 & 0.225 \\
 & $\hat{\omega}$ & -0.027 & 0.103 & -0.132 & 0.202 & -0.156 & 0.173 & -0.006 & 0.117 \\
 & $\hat{\alpha}$ & 0.606 & 4.812 & $>$100 & $>$100 & -3.042 & 3.072 & -0.414 & 1.627 \\
\addlinespace
\midrule
\bottomrule
\end{tabular}
\end{table}

\begin{table}[!htb]
\centering
\caption{Median and IQR of parameter estimates for MLE, MOM, MWM, and GRID-MOM based on $1{,}000$ replications with sample size $n = 500$ under truncation rate $\tau = 0.1$. The true parameters are $\xi_0 = 0$, $\omega_0 = 1$, and $\alpha_0 \in \{1, 2, 4\}$.}
\label{tab:table3}
\begin{tabular}{llcccccccc}
\toprule
Truncation & & \multicolumn{2}{c}{MLE} & \multicolumn{2}{c}{MOM} & \multicolumn{2}{c}{MWM} & \multicolumn{2}{c}{GRID-MOM} \\
\cmidrule(lr){3-4} \cmidrule(lr){5-6} \cmidrule(lr){7-8} \cmidrule(lr){9-10}
 &  & Median & IQR & Median & IQR & Median & IQR & Median & IQR \\
\midrule
\multicolumn{10}{l}{$\alpha_0 = 1$} \\
Left & $\hat{\xi}$ & 0.031 & 0.540 & 0.344 & 0.909 & 0.074 & 0.121 & 0.028 & 0.959 \\
 & $\hat{\omega}$ & 0.982 & 0.171 & 0.987 & 0.152 & 0.958 & 0.064 & 1.038 & 0.145 \\
 & $\hat{\alpha}$ & 0.971 & 1.413 & 0.151 & 1.638 & 0.774 & 0.054 & 0.975 & 1.975 \\
\addlinespace
Right & $\hat{\xi}$ & 0.007 & 0.664 & -0.000 & 1.025 & 0.005 & 0.631 & 0.008 & 1.027 \\
 & $\hat{\omega}$ & 0.991 & 0.387 & 1.009 & 0.297 & 0.988 & 0.393 & 1.013 & 0.300 \\
 & $\hat{\alpha}$ & 0.994 & 1.581 & 0.999 & 2.157 & 1.005 & 1.508 & 1.000 & 2.175 \\
\addlinespace
Double & $\hat{\xi}$ & 0.012 & 0.703 & 0.060 & 1.272 & 0.055 & 0.548 & 0.011 & 1.286 \\
 & $\hat{\omega}$ & 0.995 & 0.347 & 1.098 & 0.271 & 0.970 & 0.308 & 1.118 & 0.262 \\
 & $\hat{\alpha}$ & 0.997 & 1.745 & 0.907 & 2.773 & 0.842 & 1.446 & 1.000 & 2.825 \\
\addlinespace
\midrule
\multicolumn{10}{l}{$\alpha_0 = 2$} \\
Left & $\hat{\xi}$ & 0.018 & 0.117 & 0.463 & 0.496 & 0.047 & 0.185 & 0.015 & 0.118 \\
 & $\hat{\omega}$ & 0.993 & 0.097 & 0.894 & 0.133 & 0.971 & 0.120 & 1.002 & 0.091 \\
 & $\hat{\alpha}$ & 2.013 & 1.436 & 0.003 & 1.260 & 1.654 & 1.471 & 2.025 & 1.425 \\
\addlinespace
Right & $\hat{\xi}$ & 0.004 & 0.155 & 0.007 & 0.154 & 0.011 & 0.162 & 0.003 & 0.154 \\
 & $\hat{\omega}$ & 0.990 & 0.233 & 0.982 & 0.232 & 0.977 & 0.242 & 0.994 & 0.233 \\
 & $\hat{\alpha}$ & 1.973 & 0.962 & 1.984 & 0.936 & 1.943 & 0.975 & 1.975 & 0.956 \\
\addlinespace
Double & $\hat{\xi}$ & 0.003 & 0.164 & 0.122 & 0.603 & 0.070 & 0.221 & 0.002 & 0.167 \\
 & $\hat{\omega}$ & 0.999 & 0.185 & 0.911 & 0.272 & 0.935 & 0.184 & 1.011 & 0.171 \\
 & $\hat{\alpha}$ & 2.069 & 1.390 & 1.291 & 2.250 & 1.596 & 1.228 & 2.075 & 1.425 \\
\addlinespace
\midrule
\multicolumn{10}{l}{$\alpha_0 = 4$} \\
Left & $\hat{\xi}$ & 0.052 & 0.107 & 0.235 & 0.470 & -0.036 & 0.191 & 0.006 & 0.114 \\
 & $\hat{\omega}$ & 0.973 & 0.076 & 0.949 & 0.150 & 0.984 & 0.105 & 1.005 & 0.085 \\
 & $\hat{\alpha}$ & 3.085 & 4.135 & 0.060 & 2.997 & 1.658 & 1.732 & 3.575 & 2.925 \\
\addlinespace
Right & $\hat{\xi}$ & -0.001 & 0.059 & -0.001 & 0.070 & 0.004 & 0.066 & -0.003 & 0.053 \\
 & $\hat{\omega}$ & 0.997 & 0.141 & 1.001 & 0.154 & 0.990 & 0.147 & 0.998 & 0.139 \\
 & $\hat{\alpha}$ & 4.068 & 1.241 & 4.057 & 1.613 & 3.908 & 1.265 & 4.050 & 1.225 \\
\addlinespace
Double & $\hat{\xi}$ & 0.016 & 0.070 & 0.415 & 0.539 & 0.136 & 0.128 & 0.008 & 0.083 \\
 & $\hat{\omega}$ & 0.982 & 0.115 & 0.847 & 0.230 & 0.834 & 0.098 & 0.990 & 0.118 \\
 & $\hat{\alpha}$ & 3.898 & 2.682 & 0.059 & 3.137 & 0.842 & 0.042 & 3.938 & 2.250 \\
\addlinespace
\midrule
\bottomrule
\end{tabular}
\end{table}

\paragraph{Result for truncation rate $\tau = 0.2$}
Table~\ref{tab:table3} reports the empirical bias and RMSE, from which several patterns emerge. 
As the truncation rate increases to $\tau = 0.2$, the performance of all methods deteriorates relative to $\tau = 0.1$, with larger bias and RMSE observed across most settings. 
MOM remains uniformly dominated by MWM across configurations. 
For MLE, numerical instability becomes more pronounced at $\tau = 0.2$. Whereas bias and RMSE exceeding 100 occur in only a single setting when $\tau = 0.1$, such extreme behavior appears in multiple simulation settings as the truncation rate increases. 
The relative performance of the competing methods depends on the truncation direction and the value of $\alpha_0$. 
MLE achieves superior performance only when $\alpha_0 = 1$ under left truncation, whereas MWM performs best when $\alpha_0 = 1$ under double truncation. 
Apart from these limited cases, estimation accuracy deteriorates substantially under left and double truncations across all methods. 
Under right truncation, however, both MWM and GRID-MOM exhibit consistently competitive performance across different values of $\alpha_0$ and clearly outperform MLE, particularly in estimating the shape parameter.

Table~\ref{tab:table4} reports the median and IQR of the parameter estimates, from which several patterns emerge. 
First, consistent with Table~\ref{tab:table3}, MWM dominates MOM; accordingly, subsequent comparisons focus on MLE, MWM, and GRID-MOM. 
Second, when $\alpha_0 = 1$ under right truncation, MLE and MWM exhibit similar performance and generally outperform GRID-MOM; although the median of GRID-MOM remains close to the true parameter values, its IQR for $\hat{\xi}$ and $\hat{\alpha}$ is slightly larger. Under left and double truncations, MLE achieves the best performance.
Third, when $\alpha_0 = 2$, performance varies across truncation directions; 
under left truncation, GRID-MOM outperforms both MLE and MWM; in this setting, the medians of the shape parameter for MLE and MWM tend to be slightly smaller than the true value. 
Under right truncation, the three methods exhibit similar performance. 
Under double truncation, MLE and GRID-MOM perform comparably and dominate MWM; notably, the median estimate of the shape parameter under MWM is substantially smaller than the true value.
Finally, when $\alpha_0 = 4$, the median of the shape parameter estimates deviates substantially from the true value under left truncation for all methods.
Accordingly, we focus the comparison on right and double truncation, where GRID-MOM exhibits performance comparable to, or better than, the competing methods.

\begin{table}[!htb]
\centering
\caption{Bias and RMSE of parameter estimates for MLE, MOM, MWM, and GRID-MOM based on $1{,}000$ replications with sample size $n = 500$ under truncation rate $\tau = 0.2$. The true parameters are $\xi_0 = 0$, $\omega_0 = 1$, and $\alpha_0 \in \{1, 2, 4\}$.}
\label{tab:table2}
\begin{tabular}{llcccccccc}
\toprule
Truncation &  & \multicolumn{2}{c}{MLE} & \multicolumn{2}{c}{MOM} & \multicolumn{2}{c}{MWM} & \multicolumn{2}{c}{GRID-MOM} \\
\cmidrule(lr){3-4} \cmidrule(lr){5-6} \cmidrule(lr){7-8} \cmidrule(lr){9-10}
 &  & Bias & RMSE & Bias & RMSE & Bias & RMSE & Bias & RMSE \\
\midrule
\multicolumn{10}{l}{$\alpha_0 = 1$} \\
Left & $\hat{\xi}$ & 0.172 & 0.331 & 0.586 & 0.762 & 0.063 & 0.177 & 0.408 & 0.725 \\
 & $\hat{\omega}$ & -0.017 & 0.206 & 0.258 & 1.107 & -0.011 & 0.084 & 0.246 & 0.612 \\
 & $\hat{\alpha}$ & 0.034 & 1.306 & $>$100 & $>$100 & 0.256 & 0.907 & -0.647 & 2.158 \\
\addlinespace
Right & $\hat{\xi}$ & 0.098 & 0.402 & 0.240 & 0.607 & 0.074 & 0.359 & 0.246 & 0.614 \\
 & $\hat{\omega}$ & 0.481 & 5.200 & 0.484 & 4.103 & 0.201 & 0.709 & 0.220 & 0.542 \\
 & $\hat{\alpha}$ & 0.838 & 11.373 & 0.432 & 9.268 & 0.314 & 1.670 & -0.164 & 1.808 \\
\addlinespace
Double & $\hat{\xi}$ & 0.145 & 0.443 & 0.545 & 0.808 & 0.075 & 0.223 & 0.418 & 0.828 \\
 & $\hat{\omega}$ & 0.095 & 0.317 & 0.128 & 0.298 & 0.002 & 0.176 & 0.272 & 0.393 \\
 & $\hat{\alpha}$ & 0.379 & 2.267 & -0.929 & 2.422 & -0.034 & 0.720 & -0.599 & 2.622 \\
\addlinespace
\midrule
\multicolumn{10}{l}{$\alpha_0 = 2$} \\
Left & $\hat{\xi}$ & 0.142 & 0.254 & 0.364 & 0.525 & 0.058 & 0.158 & 0.245 & 0.515 \\
 & $\hat{\omega}$ & 0.080 & 1.875 & 0.156 & 1.221 & -0.032 & 0.082 & 0.173 & 0.628 \\
 & $\hat{\alpha}$ & $>$100 & $>$100 & $>$100 & $>$100 & -0.276 & 1.105 & -0.250 & 2.505 \\
\addlinespace
Right & $\hat{\xi}$ & 0.043 & 0.203 & 0.065 & 0.266 & 0.047 & 0.192 & 0.063 & 0.262 \\
 & $\hat{\omega}$ & 0.122 & 1.568 & 0.085 & 0.840 & 0.033 & 0.332 & 0.050 & 0.319 \\
 & $\hat{\alpha}$ & 0.394 & 4.589 & 0.226 & 2.793 & 0.109 & 1.234 & 0.096 & 1.316 \\
\addlinespace
Double & $\hat{\xi}$ & 0.170 & 0.347 & 0.543 & 0.696 & 0.165 & 0.191 & 0.311 & 0.623 \\
 & $\hat{\omega}$ & -0.031 & 0.241 & -0.080 & 0.302 & -0.156 & 0.184 & 0.084 & 0.263 \\
 & $\hat{\alpha}$ & 0.771 & 14.697 & $>$100 & $>$100 & -1.090 & 1.117 & -0.540 & 2.604 \\
\addlinespace
\midrule
\multicolumn{10}{l}{$\alpha_0 = 4$} \\
Left & $\hat{\xi}$ & 0.001 & 0.266 & 0.160 & 0.475 & -0.177 & 0.300 & 0.171 & 0.565 \\
 & $\hat{\omega}$ & 0.107 & 1.540 & 0.680 & 2.682 & 0.022 & 0.095 & 0.686 & 1.455 \\
 & $\hat{\alpha}$ & $>$100 & $>$100 & $>$100 & $>$100 & -2.884 & 2.988 & -2.960 & 4.508 \\
\addlinespace
Right & $\hat{\xi}$ & 0.004 & 0.052 & 0.011 & 0.129 & 0.013 & 0.055 & 0.008 & 0.047 \\
 & $\hat{\omega}$ & 0.026 & 0.206 & 0.032 & 0.260 & -0.002 & 0.171 & 0.006 & 0.151 \\
 & $\hat{\alpha}$ & 0.260 & 1.402 & 0.388 & 2.419 & -0.048 & 1.089 & -0.004 & 0.857 \\
\addlinespace
Double & $\hat{\xi}$ & 0.081 & 0.219 & 0.374 & 0.546 & 0.017 & 0.194 & 0.144 & 0.467 \\
 & $\hat{\omega}$ & 0.025 & 2.326 & 0.525 & 5.928 & -0.119 & 0.181 & 0.135 & 0.508 \\
 & $\hat{\alpha}$ & $>$100 & $>$100 & $>$100 & $>$100 & -3.162 & 3.164 & -1.234 & 3.083 \\
\addlinespace
\midrule
\bottomrule
\end{tabular}
\end{table}

\begin{table}[!htb]
\centering
\caption{Median and IQR of parameter estimates for MLE, MOM, MWM, and GRID-MOM based on $1{,}000$ replications with sample size $n = 500$ under truncation rate $\tau = 0.2$. The true parameters are $\xi_0 = 0$, $\omega_0 = 1$, and $\alpha_0 \in \{1, 2, 4\}$.}
\label{tab:table4}
\begin{tabular}{llcccccccc}
\toprule
Truncation & & \multicolumn{2}{c}{MLE} & \multicolumn{2}{c}{MOM} & \multicolumn{2}{c}{MWM} & \multicolumn{2}{c}{GRID-MOM} \\
\cmidrule(lr){3-4} \cmidrule(lr){5-6} \cmidrule(lr){7-8} \cmidrule(lr){9-10}
 &  & Median & IQR & Median & IQR & Median & IQR & Median & IQR \\
\midrule
\multicolumn{10}{l}{$\alpha_0 = 1$} \\
Left & $\hat{\xi}$ & 0.057 & 0.480 & 0.521 & 0.759 & 0.019 & 0.245 & 0.053 & 1.099 \\
 & $\hat{\omega}$ & 0.978 & 0.150 & 0.957 & 0.215 & 0.998 & 0.137 & 1.053 & 0.172 \\
 & $\hat{\alpha}$ & 0.914 & 1.634 & -0.067 & 1.385 & 0.808 & 1.732 & 0.938 & 2.706 \\
\addlinespace
Right & $\hat{\xi}$ & 0.016 & 0.722 & 0.011 & 1.123 & 0.021 & 0.677 & 0.015 & 1.130 \\
 & $\hat{\omega}$ & 0.989 & 0.591 & 1.002 & 0.517 & 0.983 & 0.576 & 1.009 & 0.503 \\
 & $\hat{\alpha}$ & 1.026 & 1.990 & 1.006 & 2.717 & 0.982 & 1.891 & 1.025 & 2.800 \\
\addlinespace
Double & $\hat{\xi}$ & 0.018 & 0.740 & 0.531 & 1.247 & 0.072 & 0.355 & 0.013 & 1.448 \\
 & $\hat{\omega}$ & 1.047 & 0.487 & 1.093 & 0.398 & 0.970 & 0.261 & 1.247 & 0.357 \\
 & $\hat{\alpha}$ & 1.046 & 2.300 & -0.001 & 2.768 & 0.819 & 1.018 & 1.050 & 4.075 \\
\addlinespace
\midrule
\multicolumn{10}{l}{$\alpha_0 = 2$} \\
Left & $\hat{\xi}$ & 0.109 & 0.170 & 0.320 & 0.472 & 0.058 & 0.168 & 0.081 & 0.166 \\
 & $\hat{\omega}$ & 0.963 & 0.092 & 0.930 & 0.138 & 0.970 & 0.095 & 0.996 & 0.102 \\
 & $\hat{\alpha}$ & 1.834 & 2.447 & 0.093 & 2.625 & 1.828 & 1.870 & 2.013 & 2.506 \\
\addlinespace
Right & $\hat{\xi}$ & -0.007 & 0.191 & -0.004 & 0.191 & -0.000 & 0.194 & -0.007 & 0.192 \\
 & $\hat{\omega}$ & 0.997 & 0.392 & 0.996 & 0.397 & 0.988 & 0.385 & 1.001 & 0.396 \\
 & $\hat{\alpha}$ & 2.004 & 1.432 & 2.010 & 1.449 & 1.969 & 1.423 & 2.013 & 1.431 \\
\addlinespace
Double & $\hat{\xi}$ & 0.039 & 0.493 & 0.581 & 0.584 & 0.175 & 0.131 & 0.031 & 0.721 \\
 & $\hat{\omega}$ & 0.966 & 0.322 & 0.840 & 0.305 & 0.832 & 0.127 & 1.050 & 0.265 \\
 & $\hat{\alpha}$ & 1.935 & 3.076 & -0.003 & 1.538 & 0.858 & 0.043 & 1.975 & 3.375 \\
\addlinespace
\midrule
\multicolumn{10}{l}{$\alpha_0 = 4$} \\
Left & $\hat{\xi}$ & 0.009 & 0.347 & 0.148 & 0.547 & -0.143 & 0.292 & 0.022 & 0.790 \\
 & $\hat{\omega}$ & 0.983 & 0.117 & 1.005 & 0.190 & 1.017 & 0.112 & 1.054 & 0.296 \\
 & $\hat{\alpha}$ & 1.105 & 2.554 & 1.013 & 2.475 & 0.808 & 1.495 & 1.988 & 4.606 \\
\addlinespace
Right & $\hat{\xi}$ & -0.000 & 0.066 & 0.003 & 0.075 & 0.010 & 0.068 & 0.001 & 0.059 \\
 & $\hat{\omega}$ & 0.999 & 0.229 & 0.998 & 0.245 & 0.984 & 0.217 & 1.003 & 0.219 \\
 & $\hat{\alpha}$ & 4.051 & 1.578 & 3.994 & 1.870 & 3.826 & 1.416 & 4.075 & 1.575 \\
\addlinespace
Double & $\hat{\xi}$ & 0.056 & 0.099 & 0.362 & 0.497 & 0.051 & 0.197 & 0.011 & 0.133 \\
 & $\hat{\omega}$ & 0.952 & 0.155 & 0.895 & 0.260 & 0.858 & 0.157 & 1.012 & 0.203 \\
 & $\hat{\alpha}$ & 3.451 & 4.346 & 0.085 & 4.251 & 0.840 & 0.064 & 3.612 & 3.100 \\
\addlinespace
\midrule
\bottomrule
\end{tabular}
\end{table}

\subsection{Comparison with GRID-MLE}\label{sim_2}
In this section, we examine a profile-likelihood-based variant of GRID-MOM. 
For convenience, we refer to this procedure as GRID-MLE and compare its performance with that of GRID-MOM. Like GRID-MOM, GRID-MLE employs a grid over the shape parameter $\alpha$; however, for each candidate value of $\alpha$, the remaining parameters $(\xi,\omega)$ are obtained by maximizing the corresponding profile likelihood rather than by solving moment equations.

\begin{figure}[!htb]
    \centering
    \includegraphics[width=0.9\linewidth]{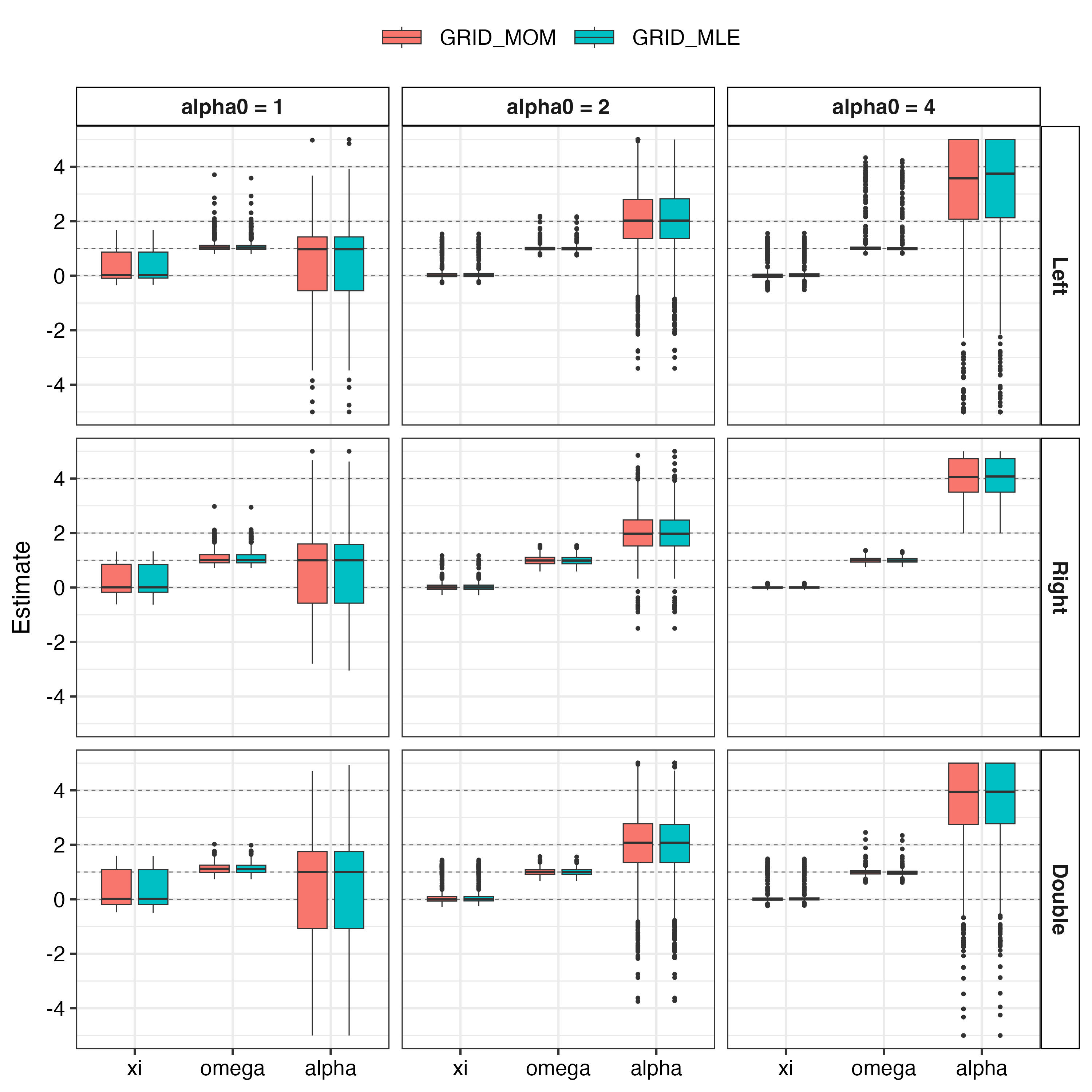}
    \caption{Comparison of GRID-MOM (red boxplots) and GRID-MLE (green boxplots) under truncation rate $\tau = 0.1$. Rows correspond to truncation direction (Left, Right, Double), and columns correspond to the true shape parameter $\alpha_0 \in \{1,2,4\}$. Across all simulation settings, the true location and scale parameters are fixed at $\xi_0 = 0$ and $\omega_0 = 1$, and the sample size is set to $n = 500$. Within each panel, the parameter estimates $(\hat{\xi}, \hat{\omega}, \hat{\alpha})$ based on 1,000 Monte Carlo replications are summarized using boxplots. The horizontal dashed lines are reference lines at 0, 1, 2, and 4.}
    \label{fig:GRID_MOM_MLE_tau=0.1}
\end{figure}

\begin{figure}[!htb]
    \centering
    \includegraphics[width=0.9\linewidth]{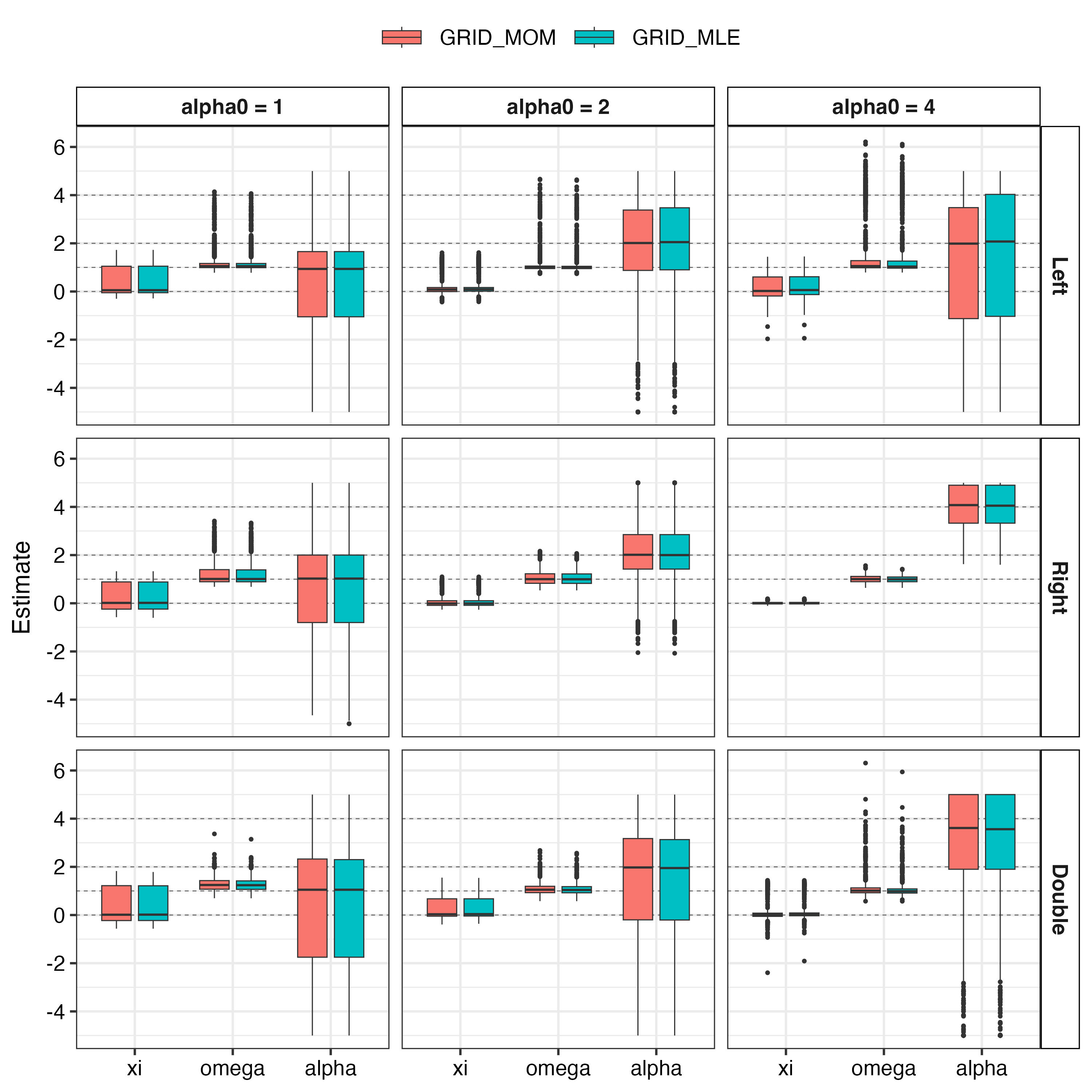}
    \caption{Comparison of GRID-MOM (red boxplots) and GRID-MLE (green boxplots) under truncation rate $\tau = 0.2$. The layout and all other simulation settings are identical to those in Figure~\ref{fig:GRID_MOM_MLE_tau=0.1}.}
    \label{fig:GRID_MOM_MLE_tau=0.2}
\end{figure}

\begin{figure}[!htb]
    \centering
    \begin{subfigure}{0.48\linewidth}
        \centering
        \includegraphics[width=\linewidth]{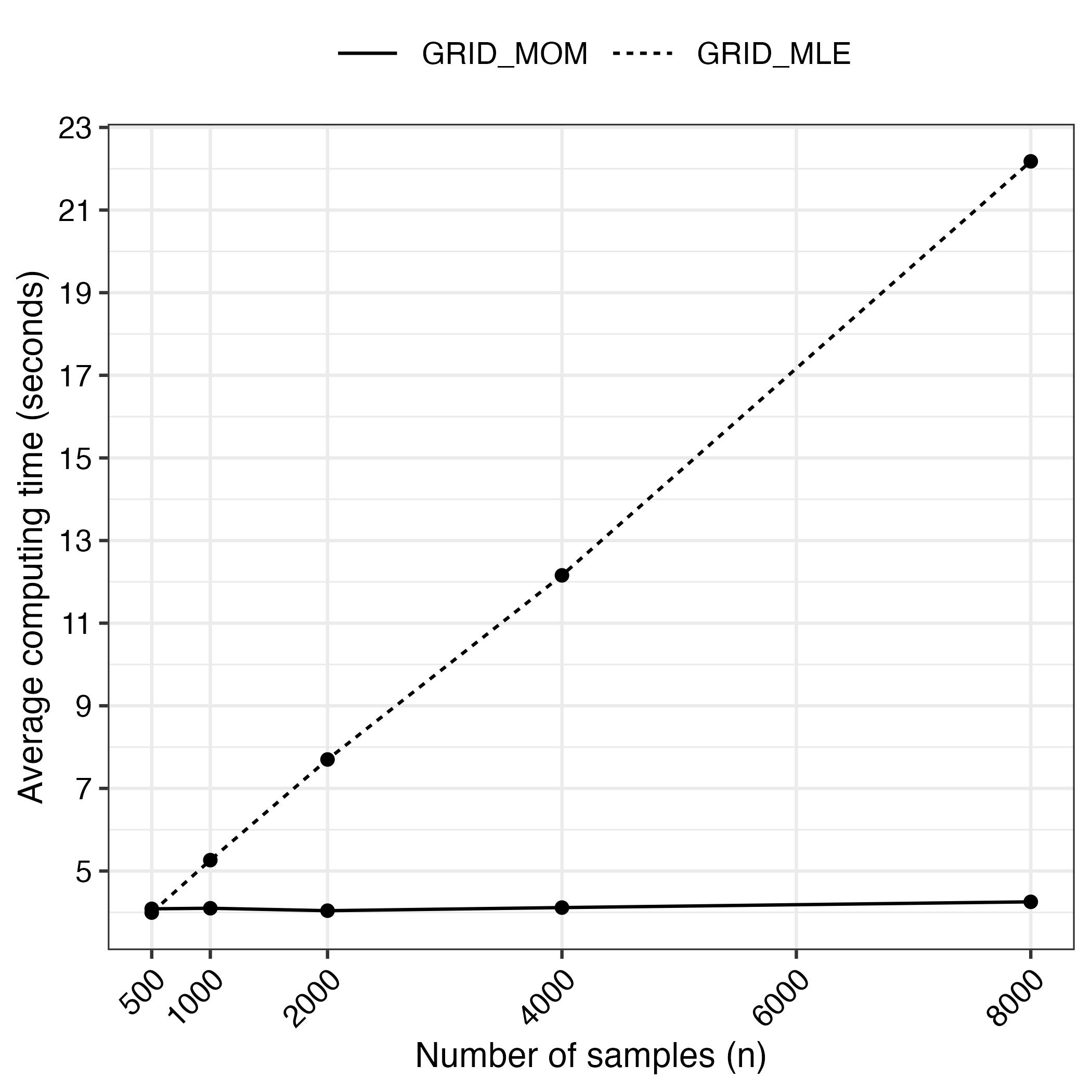}
        \caption{Avg. computing time vs. sample size}
        \label{fig:cost_n}
    \end{subfigure}
    \hspace{0.01\linewidth}
    \begin{subfigure}{0.48\linewidth}
        \centering
        \includegraphics[width=\linewidth]{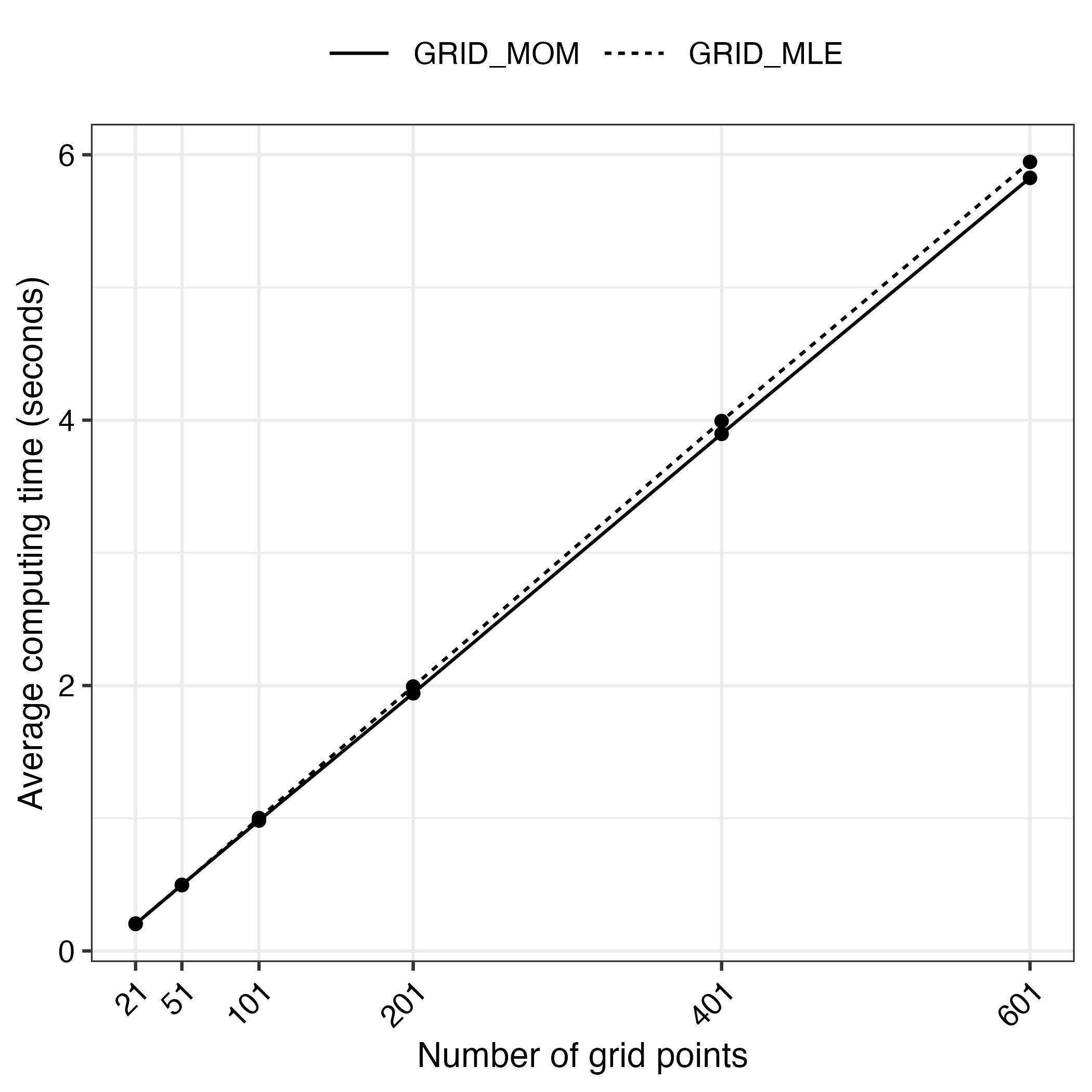}
        \caption{Avg. computing time vs. number of grid points}
        \label{fig:cost_grid}
    \end{subfigure}
    \caption{Computational cost of GRID-MOM and GRID-MLE as functions of sample size (left panel) and the number of grid points (right panel). In the left panel, the reported average computing times are based on 50 Monte Carlo replications with the number of grid points fixed at 401. In the right panel, the averages are based on 100 Monte Carlo replications with the sample size fixed at $n = 500$.}
    \label{fig:computing_cost}
\end{figure}

Figures~\ref{fig:GRID_MOM_MLE_tau=0.1}--\ref{fig:GRID_MOM_MLE_tau=0.2} compare the estimates from GRID-MOM and GRID-MLE under the same simulation settings as in Section~\ref{sim_1}. We observe that both methods exhibit nearly identical performance across all simulation scenarios considered. 
Figure~\ref{fig:computing_cost} reports the average computing time as functions of sample size and the number of grid points. The results indicate that the computational cost of GRID-MOM is consistently lower than that of GRID-MLE. In particular, as the sample size increases, GRID-MLE requires substantially more computation time than GRID-MOM, and the gap becomes more pronounced. 
In summary, GRID-MOM achieves comparable estimation performance while requiring substantially less computational effort than GRID-MLE. Therefore, we conclude that GRID-MOM provides a computationally efficient alternative to the profile likelihood approach.

\section{Real Data Example}\label{sec:realdata}
First, we demonstrate the practical performance of GRID-MOM using TCGA high-grade serous ovarian carcinoma (HGSC) phosphoproteomics data. The dataset consists of normalized phosphorylation measurements for 5{,}746 sites across 67 tumor samples, classified into five proteomic subtypes (A–E) \citep{zhang2016integrated, seo2025multiple}. 

Among the 20 possible pairwise subtype comparisons, we focus on a representative case comparing subtype D (Mesenchymal) to subtype B (Immunoreactive). Our objective is to identify phosphorylation sites whose mean phosphorylation levels are higher in subtype D than in subtype B. Solving this one-sided multiple testing problem requires accurate estimation of the null density of the test statistics. Under the zero assumption \citep{efron2004large} and assuming that the null distribution belongs to the skew-normal family, the problem reduces to parameter estimation for a truncated skew-normal distribution based on truncated test statistics. 

\begin{figure}[htb!]
    \centering
    \includegraphics[width=0.7\linewidth]{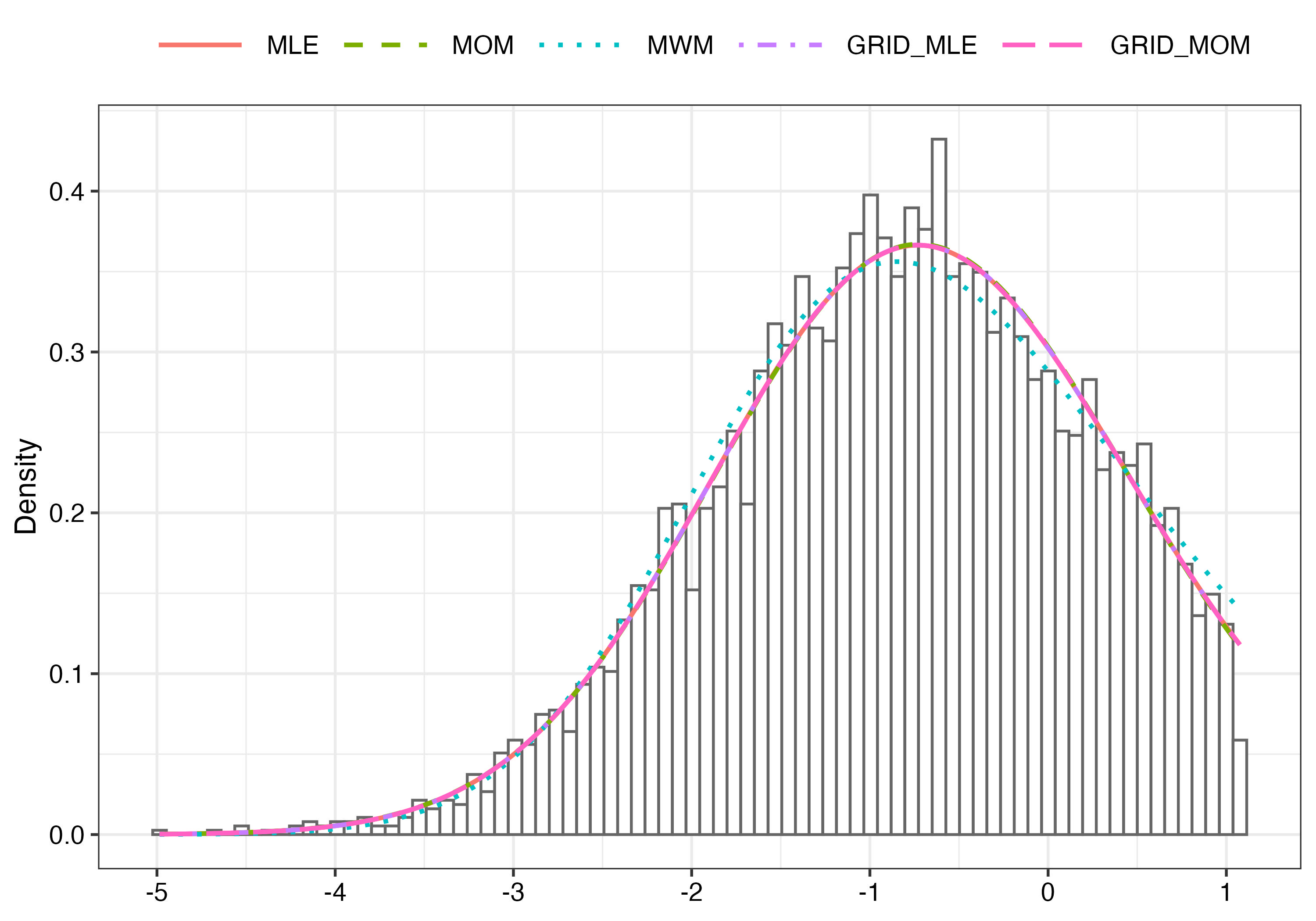}
    \caption{Histogram of the test statistics from the phosphoproteomics data, overlaid with the estimated truncated densities obtained by MLE (red solid line), MOM (green dashed line), MWM (blue dotted line), GRID-MLE (purple dot-dashed line), and GRID-MOM (pink long-dashed line). The fitted density curves from MLE, MOM, GRID-MLE, and GRID-MOM are nearly indistinguishable and overlap almost completely.}
    \label{fig:1}
\end{figure}

For each phosphorylation site, a two-sample $t$-statistic is computed to assess elevated phosphorylation levels between the two subtypes. To implement the zero assumption, we truncate the upper 15\% of the observed $t$-statistics and fit a truncated skew-normal model to the remaining $t$-statistics using MLE, MOM, MWM, GRID-MLE, and GIRD-MOM. In contrast to the numerical studies, we adopt a multi-start strategy for MLE. Specifically, let $m$ and $s$ denote the sample mean and sample standard deviation of the observations, respectively. We generate 20 candidate starting values by sampling $(\xi, \omega, \alpha)$ uniformly from the cube $[m-1, m+1] \times [\max(s-1,1), \max(s+1,2)] \times [-5,5]$. For each starting value, we compute the corresponding MLE and select the solution with the highest log-likelihood value as the final estimate. The results are summarize in Figure~\ref{fig:1}. 

All five methods produce estimated truncated skew-normal densities that closely align with the empirical histogram. In particular, MLE, MOM, GRID-MLE, and GRID-MOM yield nearly identical fitted densities, while MWM produces a slightly different curve that remains broadly similar. 
Specifically, the estimated parameters for MLE, MOM, GRID-MLE, and GRID-MOM are $(\hat{\xi}, \hat{\omega}, \hat{\alpha}) = (-1.391, 1.379, 0.861)$, $(-1.343, 1.347, 0.785)$, $(-1.384, 1.375, 0.850)$, and $(-1.384, 1.375, 0.850)$, respectively, whereas the corresponding estimates for MWM are $(-1.836, 1.842, 1.864)$. The slight difference in the fitted density under MWM arises because its parameter estimates differ slightly from those obtained by the other methods.

As a second example, we analyze data on the number of hospital admission days for patients with dementia, measured over a one-year period. Because individual-level hospital admission days are considered sensitive personal information, direct access to such data is restricted. We therefore construct a synthetic dataset based on publicly available national health insurance statistics. The resulting dataset consists of hospital admission days for 6,000 patients with dementia, and our goal is to model the distribution of these admission days.

The synthetic dataset was constructed using aggregated statistics provided by the Health Insurance Review and Assessment Service (HIRA) in Korea, which report the number of patients and total days of care by diagnosis code, sex, and age group. Since these statistics aggregate both inpatient and outpatient services, we applied inpatient utilization ratios provided by HIRA to obtain inpatient-specific lengths of stay. Our analysis focused on dementia-related diagnoses corresponding to ICD-10 codes F00–F03.
To approximate realistic hospitalization patterns, individuals were sampled via multinomial allocation to preserve the empirical population composition, and individual lengths of stay were generated from a log-normal distribution\footnote{The log-normal distribution was chosen as it naturally accommodates the non-negativity, strong right skewness, and large variability typically observed in hospital length-of-stay data.} calibrated to the reported mean lengths of stay and a coefficient of variation obtained from a 2018 report published by NHIS Ilsan Hospital.

Since the observed admission days range from a minimum of 1 day to a maximum of 356 days, we treat the data as truncated and model the distribution using a truncated skew-normal distribution. Because the distribution of admission days is highly right-skewed, a square-root transformation is applied prior to fitting the truncated skew-normal model. As in the phosphoproteomics data analysis, we apply MLE with multiple initial values.

\begin{figure}[htb!] 
    \centering
    \includegraphics[width=0.7\linewidth]{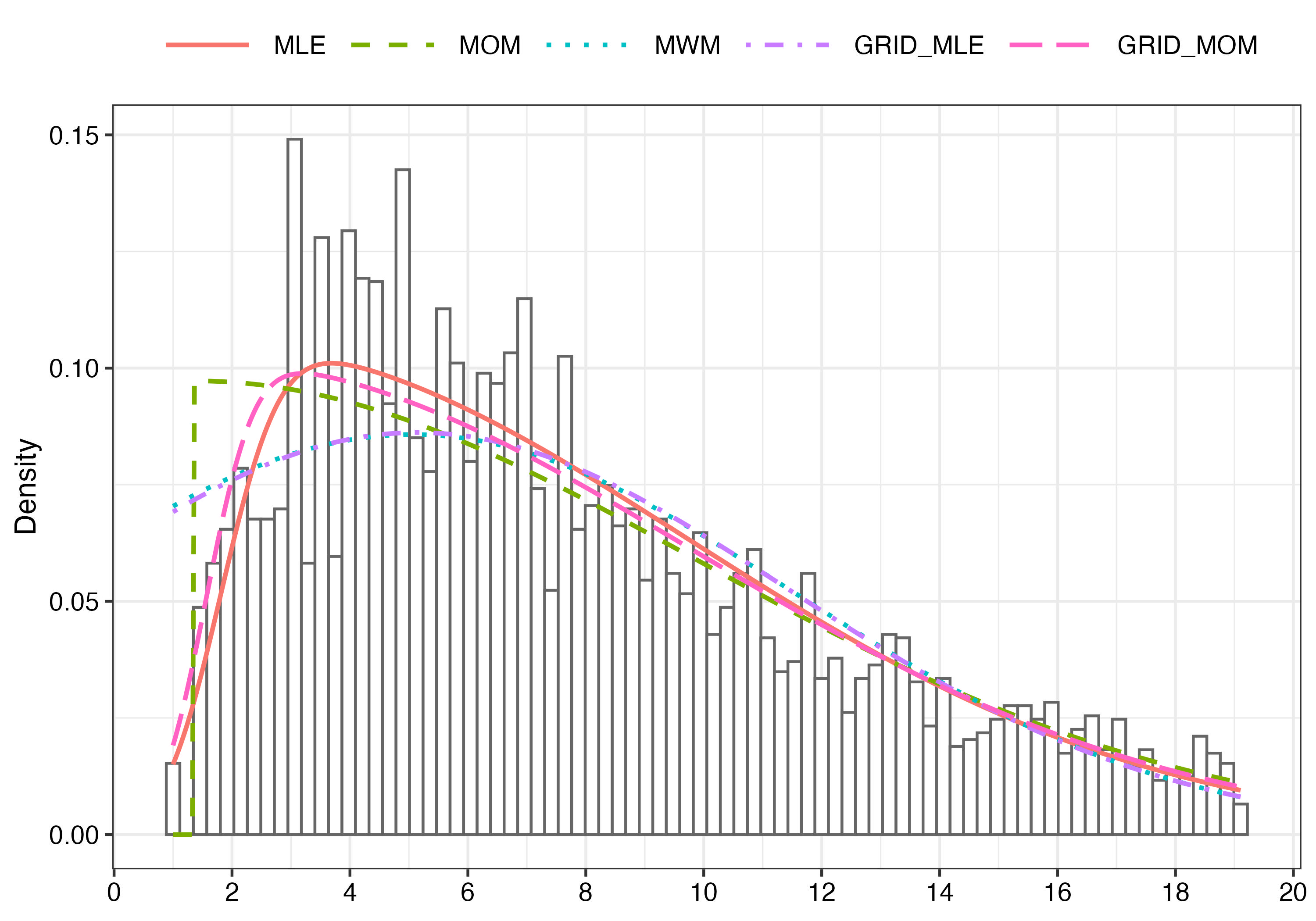}
    \caption{Histogram of the hospital admission days, overlaid with the estimated truncated densities obtained by MLE (red solid line), MOM (green dashed line), MWM (blue dotted line), GRID-MLE (purple dot-dashed line), and GRID-MOM (pink long-dashed line). The fitted density curves from MWM and GRID-MLE are nearly indistinguishable and overlap almost completely.}
    \label{fig:2}
\end{figure}

Figure~\ref{fig:2} shows the histogram of the observations overlaid with the estimated densities. Although the histogram exhibits noticeable skewness, its shape is not sufficiently smooth to be well described by the skew-normal family. Nevertheless, we employ the skew-normal distribution as a convenient parametric model for density fitting. 
The fitted densities from MLE and GRID-MOM closely align with each other, whereas MOM appears to overestimate the degree of skewness. The fitted densities from MWM and GRID-MLE are nearly identical and appear to ignore the peak around 4. 
Specifically, the MLE and GRID-MOM estimates are $(\hat{\xi}, \hat{\omega}, \hat{\alpha}) = (1.826, 7.883, 10.116)$ and $(1.563, 8.187, 12.825)$, respectively, while the MOM estimates are $(1.347, 8.520, > 100)$. The corresponding estimates for MWM and GRID-MLE are $(5.237, 6.461, -0.035)$ and $(2.162, 7.043, 0.675)$, respectively. 
The MOM estimator yields an extremely large estimate of the shape parameter, which explains the pronounced rightward skewness observed in the fitted density. In contrast, MLE and GRID-MOM produce large shape parameter estimates, whereas MWM and GRID-MLE yield relatively small ones, resulting in markedly different fitted densities. 
The discrepancy between GRID-MOM and GRID-MLE may arise from the fact that the underlying observations do not exactly follow a truncated skew-normal distribution.

\section{Conclusion}\label{sec:discussion}
We have investigated the problem of parameter estimation for the truncated skew-normal distribution and proposed a grid-based estimation method that combines moment-based estimation with likelihood evaluation. 
By decoupling the estimation of the shape parameter from that of the location and scale parameters, the proposed approach reduces the complexity of the estimation problem and improves numerical stability. 
Through an extensive numerical study, we compared the proposed estimator with MLE, MOM and MWM. The results demonstrate that the proposed method achieves competitive or superior finite-sample performance, particularly in terms of stability and accuracy for the shape parameter. 
Overall, the proposed method provides a simple framework for parameter estimation in truncated skew-normal models and offers a useful alternative to existing approaches in applications where truncation and skewness play a central role.

Throughout this paper, we have focused primarily on point estimation. For practical inference, however, quantifying uncertainty is equally important. To this end, we propose a parametric bootstrap procedure to assess the variability of the GRID-MOM estimator, following \citet{flecher2010truncated}.
Suppose that $n$ observations are available from a truncated skew-normal distribution with truncation interval $[L, U]$. Let $(\hat{\xi}, \hat{\omega}, \hat{\alpha})$ denote the GRID-MOM estimates based on the observed data. Treating these estimates as the data-generating parameters, we generate a bootstrap sample of size $n$ from the truncated skew-normal distribution with parameters $(\hat{\xi}, \hat{\omega}, \hat{\alpha})$ under the same truncation interval $[L, U]$. GRID-MOM is then reapplied to the bootstrap sample to obtain a new set of parameter estimates. Repeating this resampling-and-refitting procedure independently $B$ times yields $B$ bootstrap replicates of the estimator. The bootstrap standard errors are computed as the sample standard deviations of these $B$ replicates.

\section*{Acknowledgement}
The authors declare that there are no conflicts of interest.

\bibliographystyle{apalike}
\bibliography{ref}

\appendix
\numberwithin{figure}{section}
\numberwithin{table}{section}

\section{Additional simulation results}

\subsection{Extreme value of shape parameter}
In this section, we examine the behavior of GRID-MOM when the true shape parameter $\alpha_0$ lies outside the prespecified search range $[-a,a]$. As in the main experiments, we set $\xi_0 = 0$ and $\omega_0 = 1$, and consider an extreme shape parameter value $\alpha_0 = 10$. Under right truncation with truncation rate $\tau = 0.1$, we generate a sample of size $n = 500$. We then apply GRID-MOM while restricting the search range to $\alpha \in [-5,5]$, thereby intentionally misspecifying the parameter space.

\begin{figure}[!htb]
    \centering
    \includegraphics[width=0.7\linewidth]{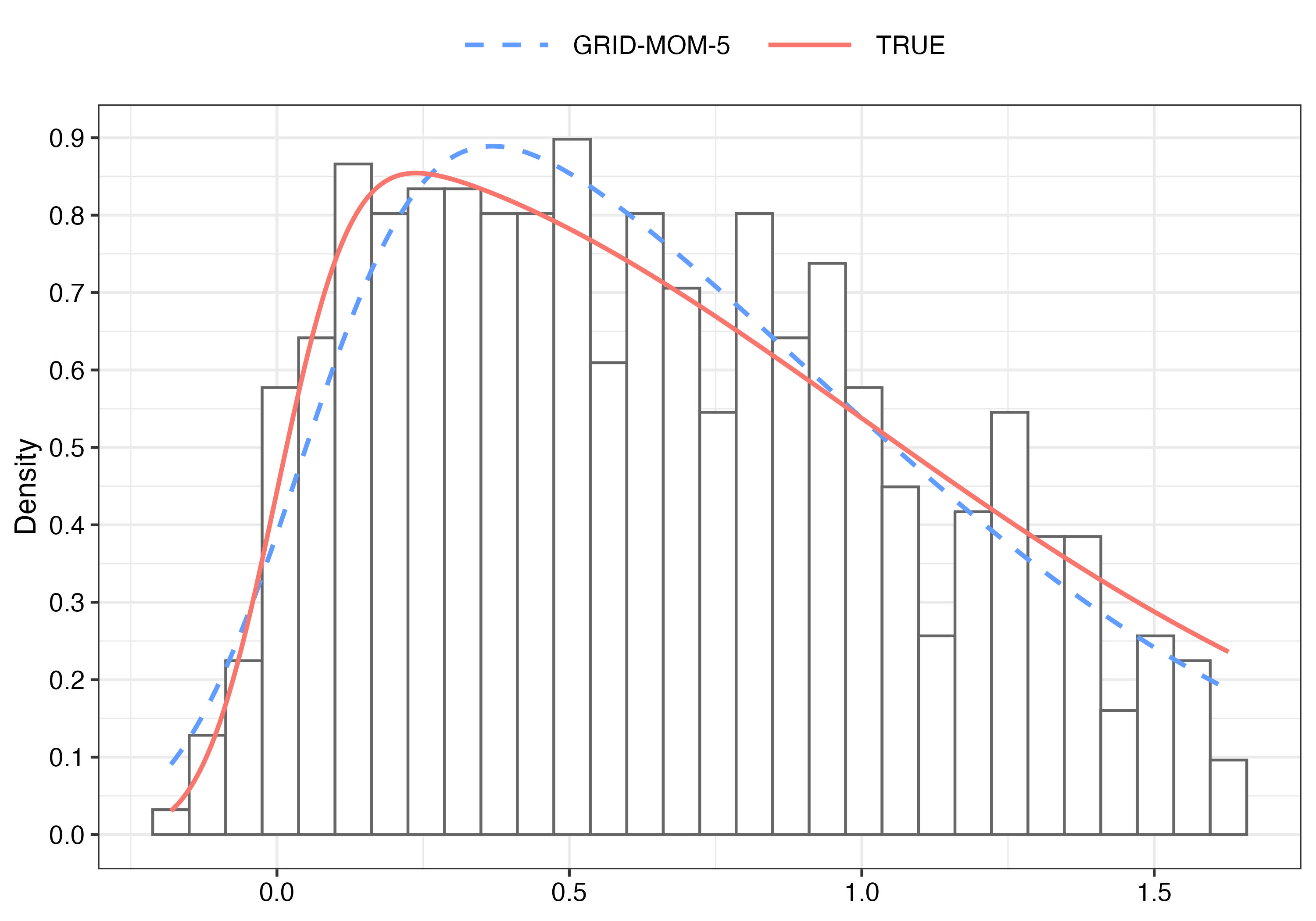}
    \caption{Histogram of $n = 500$ observations generated from a truncated skew-normal distribution with $(\xi_0, \omega_0, \alpha_0) = (0,1,10)$ under right truncation at rate $\tau = 0.1$, overlaid with the true density and GRID-MOM–based density estimate. The curve labeled TRUE represents the true truncated density. The curve labeled GRID-MOM-5 correspond to density estimates obtained by GRID-MOM under the restriction $\alpha \in [-5,5]$.}
    \label{fig:extreme_ex}
\end{figure}

The resulting estimates are $\hat{\xi} = 0.046$, $\hat{\omega} = 0.868$, and $\hat{\alpha} = 5$. Because the true value $\alpha_0 = 10$ lies outside the imposed search range, GRID-MOM cannot recover the true shape parameter, and the estimate of $\alpha$ is consequently biased. To assess the impact of this bias on the fitted density, we overlay the true and estimated densities on the empirical histogram of the observations. As shown in Figure~\ref{fig:extreme_ex}, the density implied by the GRID-MOM estimates (blue dashed line) differs from the true density (red solid line), yet remains closely aligned with the empirical histogram. This behavior arises because, for fixed $(\xi_0,\omega_0)$, skew-normal densities corresponding to sufficiently large values of $\alpha$ become nearly indistinguishable. Beyond a certain threshold---depending on $(\xi_0,\omega_0)$---further increases in $\alpha$ have only a limited effect on the overall shape of the density. 
Thus, although restricting the parameter space may prevent exact recovery of the true $\alpha$, the recommended range for $\alpha$ is typically sufficient to capture the essential features of the underlying density in practice.

When prior information on the degree of skewness is available, the search range for $\alpha$ may be expanded accordingly. In this example, extending the range to $[-15,15]$ yields the estimates $\hat{\xi} = 0.012$, $\hat{\omega} = 0.919$, and $\hat{\alpha} = 9$, which are substantially closer to the true parameter values.

\subsection{Results for small sample sizes}
In this section, we conduct an additional numerical study to evaluate and compare the performance of the methods under a reduced sample size. Specifically, we set the sample size to $n = 100$ and the truncation rate to $\tau = 0.1$, while keeping all other simulation settings identical to those in the main manuscript. The results are summarized in Tables~\ref{tab:tsn_bias_rmse}--\ref{tab:tsn_median_iqr}. 

Table~\ref{tab:tsn_bias_rmse} reports the empirical bias and RMSE, from which several patterns emerge. 
Across all simulation settings, the bias and RMSE of MLE and MOM for the shape parameter exceed 100, indicating severe instability. Moreover, for MLE, the bias and RMSE of the scale parameter also exceed 100 in several settings, further underscoring its numerical instability. For this reason, subsequent comparisons focus on MWM and GRID-MOM.
Under right truncation with $\alpha_0 \in \{2,4\}$, GRID-MOM consistently outperforms MWM. In the remaining simulation settings, however, both MWM and GRID-MOM exhibit substantial bias and/or RMSE.

Table~\ref{tab:tsn_median_iqr} presents the median and IQR of the estimators. 
Overall, MLE and GRID-MOM perform comparably to or better than MOM and MWM; accordingly, subsequent comparisons focus on MLE and GRID-MOM. 
When $\alpha_0 \in \{1,2\}$, under left truncation the median estimate of the shape parameter from MLE tends to deviate slightly more from the true value than that from GRID-MOM. Under the remaining truncation directions, MLE and GRID-MOM yield similar medians close to the true values, while the IQR of MLE is consistently smaller than that of GRID-MOM. 
When $\alpha_0 = 4$, both methods exhibit substantial deviation of the median shape estimates under left truncation. Under right and double truncation, however, GRID-MOM outperforms MLE.

\begin{table}[!htb]
\centering
\caption{Bias and RMSE of parameter estimates for MLE, MOM, MWM, and GRID-MOM based on $1{,}000$ replications with sample size $n = 100$ under truncation rate $\tau = 0.1$. The true parameters are $\xi_0 = 0$, $\omega_0 = 1$, and $\alpha_0 \in \{1, 2, 4\}$.}
\label{tab:tsn_bias_rmse}
\begin{tabular}{llcccccccc}
\toprule
Truncation &  & \multicolumn{2}{c}{MLE} & \multicolumn{2}{c}{MOM} & \multicolumn{2}{c}{MWM} & \multicolumn{2}{c}{GRID-MOM} \\
\cmidrule(lr){3-4} \cmidrule(lr){5-6} \cmidrule(lr){7-8} \cmidrule(lr){9-10}
 &  & Bias & RMSE & Bias & RMSE & Bias & RMSE & Bias & RMSE \\
\midrule
\multicolumn{10}{l}{$\alpha_0 = 1$} \\
Left & $\hat{\xi}$ & 0.199 & 0.402 & 0.690 & 0.972 & 0.063 & 0.240 & 0.566 & 0.952 \\
 & $\hat{\omega}$ & 0.100 & 4.218 & 0.687 & 3.310 & -0.018 & 0.141 & 0.395 & 0.761 \\
 & $\hat{\alpha}$ & $>$100 & $>$100 & $>$100 & $>$100 & 0.297 & 11.969 & -0.998 & 2.951 \\
\addlinespace
Right & $\hat{\xi}$ & 0.027 & 0.457 & 0.222 & 0.716 & 0.042 & 0.442 & 0.239 & 0.734 \\
 & $\hat{\omega}$ & $>$100 & $>$100 & 1.681 & 16.105 & 0.638 & 5.277 & 0.324 & 0.642 \\
 & $\hat{\alpha}$ & $>$100 & $>$100 & $>$100 & $>$100 & 1.452 & 10.130 & -0.041 & 2.517 \\
\addlinespace
Double & $\hat{\xi}$ & 0.108 & 0.443 & 0.489 & 0.913 & 0.142 & 0.408 & 0.447 & 0.931 \\
 & $\hat{\omega}$ & $>$100 & $>$100 & 0.449 & 4.333 & 0.057 & 0.336 & 0.363 & 0.591 \\
 & $\hat{\alpha}$ & $>$100 & $>$100 & $>$100 & $>$100 & 19.284 & $>$100 & -0.573 & 3.164 \\
\addlinespace
\midrule
\multicolumn{10}{l}{$\alpha_0 = 2$} \\
Left & $\hat{\xi}$ & 0.185 & 0.336 & 0.532 & 0.764 & 0.077 & 0.248 & 0.339 & 0.691 \\
 & $\hat{\omega}$ & -0.062 & 0.155 & 0.442 & 2.515 & -0.041 & 0.147 & 0.232 & 0.687 \\
 & $\hat{\alpha}$ & $>$100 & $>$100 & $>$100 & $>$100 & -0.499 & 1.068 & -0.498 & 2.964 \\
\addlinespace
Right & $\hat{\xi}$ & 0.076 & 0.306 & 0.135 & 0.428 & 0.104 & 0.305 & 0.143 & 0.445 \\
 & $\hat{\omega}$ & $>$100 & $>$100 & 0.154 & 3.009 & 0.015 & 0.458 & 0.052 & 0.351 \\
 & $\hat{\alpha}$ & $>$100 & $>$100 & $>$100 & $>$100 & 0.216 & 2.394 & 0.013 & 1.973 \\
\addlinespace
Double & $\hat{\xi}$ & 0.158 & 0.366 & 0.468 & 0.718 & 0.223 & 0.336 & 0.321 & 0.704 \\
 & $\hat{\omega}$ & 0.165 & 5.844 & 0.099 & 2.082 & -0.122 & 0.243 & 0.156 & 0.430 \\
 & $\hat{\alpha}$ & $>$100 & $>$100 & $>$100 & $>$100 & -0.627 & 4.937 & -0.453 & 3.084 \\
\addlinespace
\midrule
\multicolumn{10}{l}{$\alpha_0 = 4$} \\
Left & $\hat{\xi}$ & 0.132 & 0.373 & 0.390 & 0.708 & -0.092 & 0.387 & 0.247 & 0.680 \\
 & $\hat{\omega}$ & 0.697 & 6.509 & 1.432 & 6.452 & -0.006 & 0.170 & 0.605 & 1.404 \\
 & $\hat{\alpha}$ & $>$100 & $>$100 & $>$100 & $>$100 & -2.522 & 2.746 & -2.674 & 4.521 \\
\addlinespace
Right & $\hat{\xi}$ & 0.017 & 0.140 & 0.027 & 0.161 & 0.047 & 0.150 & 0.034 & 0.147 \\
 & $\hat{\omega}$ & $>$100 & $>$100 & 0.102 & 1.776 & -0.007 & 0.323 & -0.002 & 0.228 \\
 & $\hat{\alpha}$ & $>$100 & $>$100 & $>$100 & $>$100 & 0.471 & 5.541 & -0.132 & 1.314 \\
\addlinespace
Double & $\hat{\xi}$ & -0.196 & 4.809 & 0.419 & 0.641 & -0.044 & 1.943 & -0.595 & 11.244 \\
 & $\hat{\omega}$ & 0.547 & 9.074 & 0.460 & 4.193 & 0.007 & 1.972 & 0.306 & 2.405 \\
 & $\hat{\alpha}$ & $>$100 & $>$100 & $>$100 & $>$100 & -3.176 & 3.236 & -1.427 & 3.435 \\
\addlinespace
\midrule
\bottomrule
\end{tabular}
\end{table}

\begin{table}[!htb]
\centering
\caption{Median and IQR of parameter estimates for MLE, MOM, MWM, and GRID-MOM based on $1{,}000$ replications with sample size $n = 100$ under truncation rate $\tau = 0.1$. The true parameters are $\xi_0 = 0$, $\omega_0 = 1$, and $\alpha_0 \in \{1, 2, 4\}$.}
\label{tab:tsn_median_iqr}
\begin{tabular}{llcccccccc}
\toprule
Truncation & & \multicolumn{2}{c}{MLE} & \multicolumn{2}{c}{MOM} & \multicolumn{2}{c}{MWM} & \multicolumn{2}{c}{GRID-MOM} \\
\cmidrule(lr){3-4} \cmidrule(lr){5-6} \cmidrule(lr){7-8} \cmidrule(lr){9-10}
 &  & Median & IQR & Median & IQR & Median & IQR & Median & IQR \\
\midrule
\multicolumn{10}{l}{$\alpha_0 = 1$} \\
Left & $\hat{\xi}$ & 0.194 & 0.643 & 0.621 & 1.282 & 0.068 & 0.324 & 0.173 & 1.484 \\
 & $\hat{\omega}$ & 0.948 & 0.235 & 1.109 & 0.477 & 0.972 & 0.175 & 1.153 & 0.428 \\
 & $\hat{\alpha}$ & 0.486 & 1.789 & -0.103 & 2.847 & 0.781 & 0.715 & 0.688 & 3.900 \\
\addlinespace
Right & $\hat{\xi}$ & -0.061 & 0.839 & -0.070 & 1.322 & -0.021 & 0.823 & -0.066 & 1.361 \\
 & $\hat{\omega}$ & 1.014 & 0.678 & 1.097 & 0.540 & 1.012 & 0.662 & 1.125 & 0.562 \\
 & $\hat{\alpha}$ & 1.187 & 2.587 & 1.186 & 3.481 & 1.135 & 2.448 & 1.200 & 3.725 \\
\addlinespace
Double & $\hat{\xi}$ & 0.045 & 0.785 & 0.212 & 1.515 & 0.234 & 0.683 & 0.024 & 1.586 \\
 & $\hat{\omega}$ & 1.014 & 0.451 & 1.213 & 0.449 & 0.957 & 0.399 & 1.273 & 0.472 \\
 & $\hat{\alpha}$ & 1.023 & 2.738 & 0.433 & 4.246 & 0.294 & 1.925 & 1.075 & 4.825 \\
\addlinespace
\midrule
\multicolumn{10}{l}{$\alpha_0 = 2$} \\
Left & $\hat{\xi}$ & 0.095 & 0.423 & 0.456 & 0.763 & 0.078 & 0.279 & 0.075 & 0.773 \\
 & $\hat{\omega}$ & 0.949 & 0.200 & 0.951 & 0.313 & 0.958 & 0.188 & 1.033 & 0.234 \\
 & $\hat{\alpha}$ & 1.717 & 3.298 & -0.003 & 1.807 & 1.505 & 1.458 & 1.963 & 3.888 \\
\addlinespace
Right & $\hat{\xi}$ & 0.001 & 0.354 & 0.009 & 0.336 & 0.035 & 0.360 & -0.007 & 0.356 \\
 & $\hat{\omega}$ & 0.956 & 0.484 & 0.954 & 0.440 & 0.917 & 0.459 & 0.971 & 0.450 \\
 & $\hat{\alpha}$ & 2.066 & 2.336 & 2.023 & 2.127 & 1.842 & 1.945 & 2.100 & 2.350 \\
\addlinespace
Double & $\hat{\xi}$ & 0.034 & 0.568 & 0.397 & 0.782 & 0.253 & 0.417 & 0.020 & 0.955 \\
 & $\hat{\omega}$ & 0.952 & 0.387 & 0.943 & 0.377 & 0.848 & 0.284 & 1.090 & 0.386 \\
 & $\hat{\alpha}$ & 1.967 & 3.888 & 0.155 & 2.687 & 0.781 & 1.402 & 2.138 & 4.575 \\
\addlinespace
\midrule
\multicolumn{10}{l}{$\alpha_0 = 4$} \\
Left & $\hat{\xi}$ & 0.123 & 0.295 & 0.331 & 0.685 & -0.026 & 0.377 & 0.060 & 0.746 \\
 & $\hat{\omega}$ & 0.937 & 0.183 & 1.006 & 0.352 & 0.979 & 0.202 & 1.056 & 0.374 \\
 & $\hat{\alpha}$ & 1.384 & 4.016 & 0.001 & 2.753 & 1.488 & 1.818 & 2.350 & 6.181 \\
\addlinespace
Right & $\hat{\xi}$ & 0.001 & 0.149 & 0.008 & 0.164 & 0.029 & 0.157 & 0.007 & 0.119 \\
 & $\hat{\omega}$ & 0.981 & 0.322 & 0.973 & 0.345 & 0.943 & 0.316 & 0.979 & 0.276 \\
 & $\hat{\alpha}$ & 4.256 & 3.503 & 3.977 & 4.067 & 3.398 & 2.513 & 4.300 & 2.000 \\
\addlinespace
Double & $\hat{\xi}$ & 0.067 & 0.226 & 0.411 & 0.599 & 0.154 & 0.295 & 0.038 & 0.242 \\
 & $\hat{\omega}$ & 0.922 & 0.252 & 0.892 & 0.347 & 0.827 & 0.235 & 1.019 & 0.327 \\
 & $\hat{\alpha}$ & 3.164 & 6.169 & 0.000 & 3.208 & 0.815 & 0.212 & 3.875 & 3.431 \\
\addlinespace
\midrule
\bottomrule
\end{tabular}
\end{table}

\end{document}